\documentclass[a4,12pt,onecolumn,oneside,notitlepage,final]{article}

\date{}
\usepackage{graphicx}
\usepackage[paperwidth=595pt,paperheight=842pt,
  top=70pt,right=70pt,bottom=70pt,left=70pt]{geometry}
\usepackage{caption}
\usepackage{titlesec}
\usepackage{microtype}
\usepackage{caption}
\usepackage{subcaption}
\usepackage{comment}
\usepackage{amsmath,mathtools}
\usepackage{float}
\usepackage{lipsum}
\usepackage{wasysym}
\usepackage[symbol]{footmisc}

\titleformat*{\section}{\fontsize{14pt}{0pt}\selectfont\bfseries}
\titleformat*{\subsection}{\normalsize\bfseries}

\makeatletter
\renewenvironment{thebibliography}[1]
{\section*{\refname\@mkboth{\refname}{\refname}}%
  \list{\@biblabel{\@arabic\c@enumiv}}%
       {\settowidth\labelwidth{\@biblabel{#1}}%
        \leftmargin\labelwidth
        \advance\leftmargin\labelsep
 \setlength\itemsep{-0.5pt}%
 \setlength\baselineskip{12pt}%
        \@openbib@code
        \usecounter{enumiv}%
        \let\p@enumiv\@empty
        \renewcommand\theenumiv{\@arabic\c@enumiv}}%
  \sloppy
  \clubpenalty4000
  \@clubpenalty\clubpenalty
  \widowpenalty4000%
  \sfcode`\.\@m}
 {\def\@noitemerr
   {\@latex@warning{Empty `thebibliography' environment}}%
   \endlist}
 \makeatother

\begin{document}

\begin{flushleft}
{\fontsize{16pt}{0pt}\selectfont\textbf{Detonation attenuation and quenching in hydrogen mixtures after the interaction with cylinders}} \\
\vspace{12pt}
{\fontsize{12pt}{0pt}\selectfont { \textbf{Hongxia Yang\textsuperscript{a}\footnote[1]{E-mail address: hyang6@uottawa.ca\\ \-\ \hspace{14.5pt} Paper presented at the 34th International Symposium on Shock Waves}, Wentian Wang\textsuperscript{a}, Farzane Zangane\textsuperscript{a}, Kevin Cheevers\textsuperscript{a}, Logan Maley\textsuperscript{a,b},  Matei Radulescu\textsuperscript{a}}}} \\
{\fontsize{12pt}{0pt}\selectfont { a Dept. of Mechanical Engineering, University of Ottawa, 161 Louis-Pasteur St., Ottawa, ON, Canada}} \\
%{\fontsize{12pt}{0pt}\selectfont { \textbf{D. Four}}} \\
{\fontsize{12pt}{0pt}\selectfont { b Mechanical Engineering Technology, Okanagan College, C245 (Science Building) 1000 KLO Road, Kelowna BC V1Y 4X8, Canada}} \\
%\vspace{12pt}
%{\fontsize{12pt}{0pt}\selectfont { Hongxia Yang (hyang6@uottawa.ca)}}\\
%\vspace{6pt}
%{\fontsize{12pt}{0pt}\selectfont {\textit{Paper presented at the 34th International Symposium on Shock Waves}}}
\end{flushleft}

\noindent
\textbf{Abstract} 

The attenuation and quenching of hydrogen-oxygen detonations transmitted across a column of cylinders were investigated experimentally and analytically at sub-atmospheric pressures. Two distinct transmission regimes were observed: successful detonation transmission and complete quenching.  The transition between the two regimes was found to correlate with the ratio of inter-cylinder separation distance $b$ to a characteristic detonation scale (cell size or induction zone length) for large blockage ratios, with critical limits comparable with those previously reported for detonation diffraction from slots.  Based on available cell size measurements, the critical transmission limit was $b/\lambda=4.5\pm3$. The proposed theoretical model based on Whitham's geometric shock dynamics confirmed the equivalence between the detonation diffraction at abrupt area changes and around cylinders with large blockage ratios.  Complete quenching observed experimentally was accounted for by the weak shock strength of the transmitted shock upon detonation failure. For the tested blockage ratios, the speed of the transmitted shock ranged between 50\% to 60\% of the Chapman-Jouguet detonation speed.  This resulted in shock temperatures below the cross-over regime for hydrogen ignition, leading to very long ignition delay times even with the Mach reflection increasing the temperature. The very long ignition delay time suppressed auto-ignition, while the high isentropic exponent prevented further convective mixing required for re-initiation. This explained the fundamental difference between the arrest of hydrogen detonations and other hydrocarbons, for which transmitted fast flames punctuated by auto-ignition events were always observed.  The strength of the transmitted shock was found to be well-predicted by our previous self-similar multiple discontinuity gas-dynamic model. Over a very narrow range near the critical conditions, both hot spot re-ignition and detonation re-initiation from Mach shock reflections were observed.

\section{Introduction}
\label{subsec:1}

Understanding the behavior of detonation transmission across a bank of cylinders is critical for both detonation physics and applications related to the design of detonation arrestors. Depending on the mixture sensitivity, the detonation wave can exhibit different behaviors: it may successfully transmit, decouple during diffraction yet re-initiate from Mach reflections, quench during diffraction but experience hot-spot ignition behind the Mach shock, or undergo complete quenching following the interaction with obstacles. For most hydrocarbon mixtures, the quenched detonation can re-amplify after traversing a substantial distance past obstacles, with the turbulent reaction zone structure closely coupled with the leading shock \cite{radulescu2011mechanism,logan2015shock,saif2017chapman,yang2022experimental}. This re-amplification has been demonstrated to correlate well with the mixture sensitivity parameter $\chi$. However, preliminary observations suggested that such fast deflagration waves are not possible in hydrogen mixtures \cite{logan2015shock}, despite having a very large $\chi$ which favors the re-amplification process. Maley \cite{logan2015shock} attributed this distinctive behavior to the high isentropic exponents of hydrogen mixtures compared to other hydrocarbon mixtures, which suppresses hydrodynamic instabilities favoring turbulent mixing.  Nevertheless, their experiments also showed that in mixtures with slightly lower sensitivities, i.e. lower initial pressures, ignition was altogether quenched. In the present investigation, we aim to re-examine this problem through new experiments over a broader range, supplemented by analytical modeling for the transmitted wave strengths and resulting chemical kinetic responses. The primary objective is to determine the critical conditions for detonation transmission and analyze the role of chemical kinetics and hydrodynamic instabilities on the propensity for high-speed flames and re-amplification to detonation in hydrogen compared to other hydrocarbon mixtures.

\section{Experiments}
\label{subsec:2}

\begin{figure}
	\begin{center}
		\includegraphics[width=0.65\linewidth]{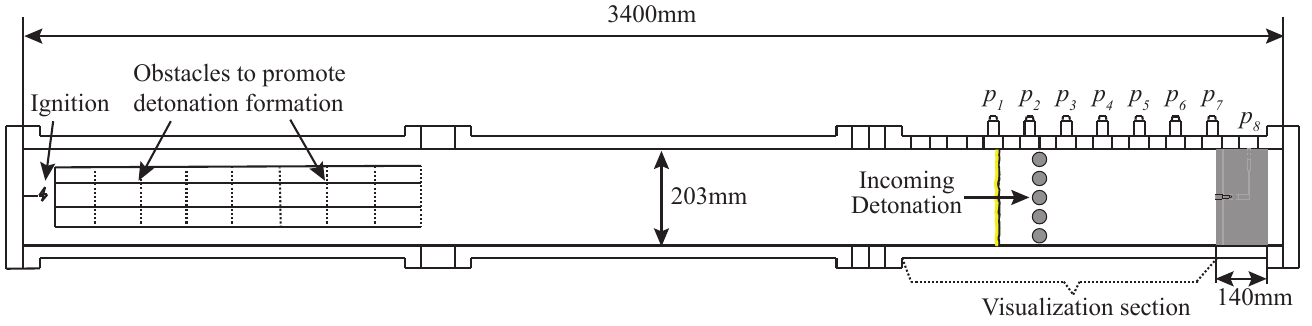}
	\end{center}
	%\captionsetup{labelformat=empty,labelsep=none}
	\caption{Schematic of the experimental setup.}
	\label{fig:exp-setup}
\end{figure}

The experiments were conducted in a shock tube with dimensions of 3400 mm $\times$ 19.1 mm $\times$ 203.2 mm, as illustrated in Fig.\ \ref{fig:exp-setup}. The last meter of the channel was equipped with glass windows allowing for visualization. A bank of cylindrical obstacles was evenly distributed at a distance of 500 mm from the end wall in the visualization section. These obstacles were intentionally selected to possess different diameters ($\diameter=$15.2 mm and 30.5 mm) and blockage ratios (BR = 60\% or 75\%) by varying the amount of cylinders, specifically 4, 5, and 10. A Z-type Schlieren system with a field of view of 318 mm in diameter was used to visualize the transmission phenomenon at various regimes. The reactive mixture studied was hydrogen-oxygen with an equivalence ratio ($\phi$) of 0.5 at an initial temperature of 294 K, with an initial pressure ranging between 9 to 20 kPa, for the overall set of experiments. The initiation of the detonation was achieved by means of a capacitor discharge at the extreme left end of the channel. The detonations reached the Chapman-Jouguet (CJ) velocity before interacting with the obstacles.  The experiments were repeated at least five times for each configuration to ensure reliability and consistency. %Further details of the experimental set-up can be found elsewhere \cite{logan2015shock,yang2022experimental}.

\begin{figure}
	\begin{center}
		\subcaptionbox{$\diameter=$ 30.5 mm, BR = 60\% \label{reinitiation} }{\includegraphics[width=0.58\linewidth]{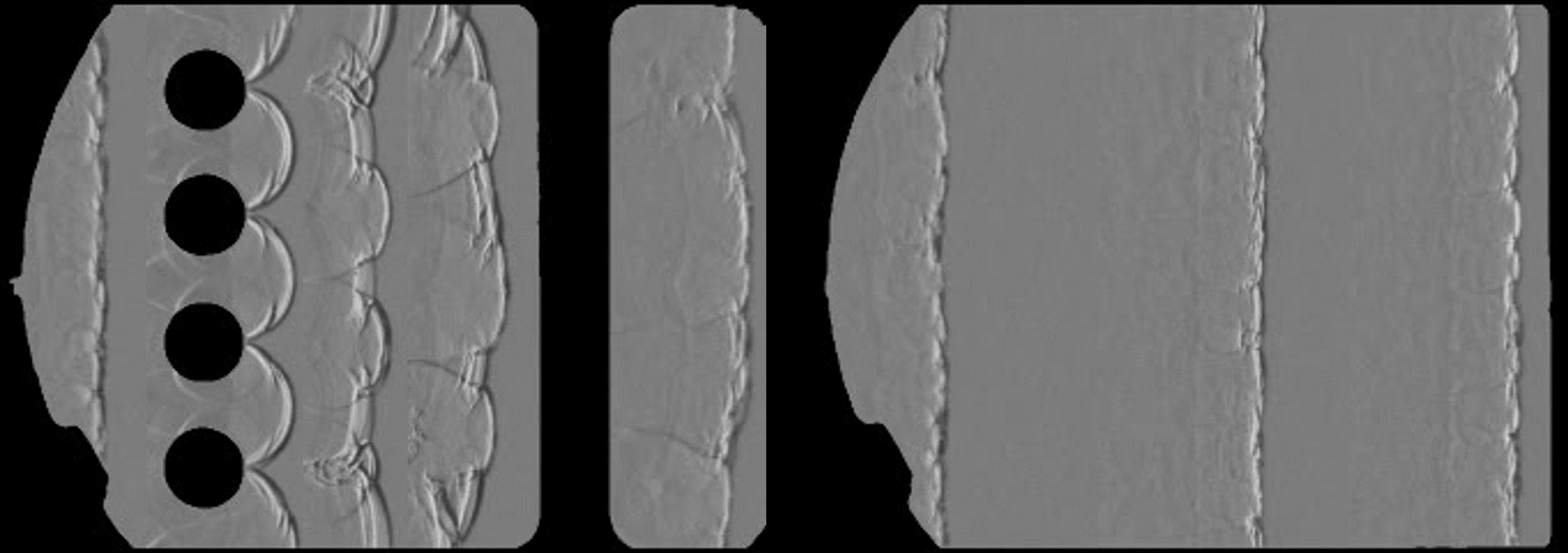}}
		\subcaptionbox{$\diameter=$ 30.5 mm, BR = 75\% \label{hotspot} }{\includegraphics[width=0.58\linewidth]{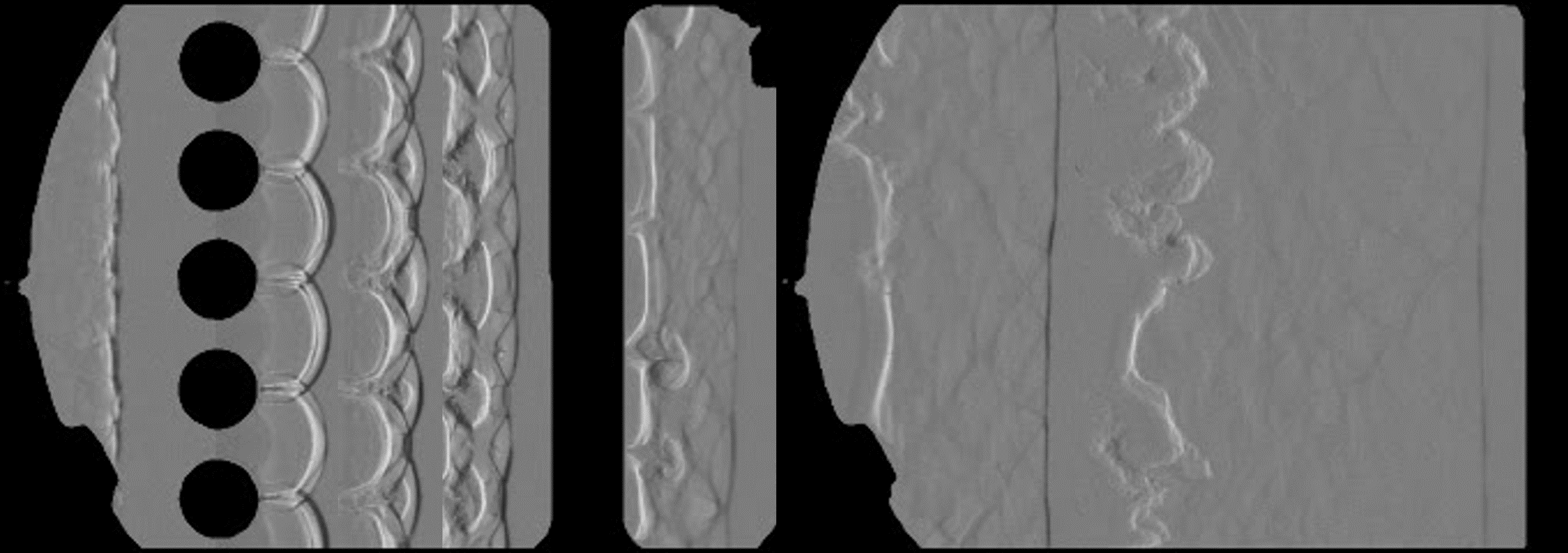}}
		\subcaptionbox{$\diameter=$ 15.25 mm, BR = 75\% \label{noinitiation} }{\includegraphics[width=0.58\linewidth]{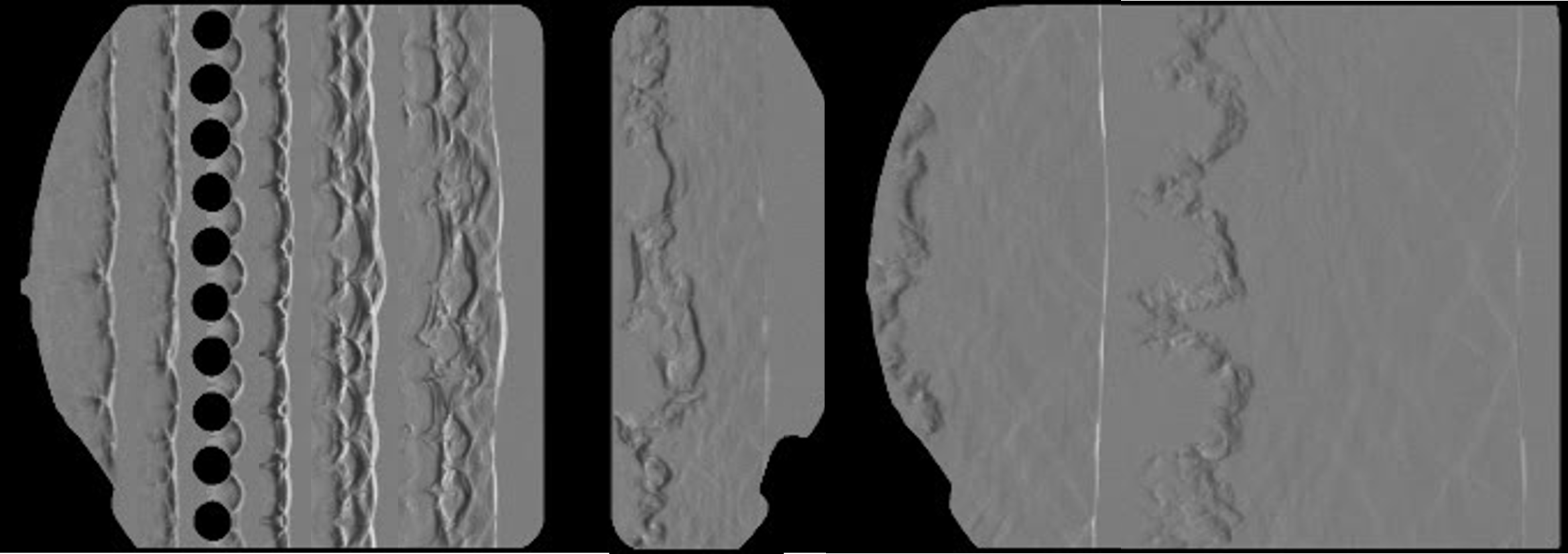}}
		%\subcaptionbox{$p_0$ = 13.8 kPa\label{det} }{\includegraphics[width=0.495\linewidth]{./figures/2H2-O2-2_00psi-Schlieren1C-detsucess-5-7-9-11nb.jpg}}
	\end{center}
	%\captionsetup{labelformat=empty,labelsep=none}
	\caption{ Composite Schlieren images illustrating (a) detonation re-initiation from first Mach reflection, (b) and (c) failed detonation with hot-spot ignition behind the first Mach reflection at an initial pressure of 9 kPa and temperature of 294 K.}
	\label{fig:exp-regimes}
\end{figure}

The experimental results have revealed consistent patterns in the behavior of hydrogen detonation transmission, irrespective of the number of obstacles or the blockage ratios employed. Once the detonation fails to re-initiate during the first Mach reflection after passing the obstacle, it continuously decays without further re-amplification. Figure \ref{fig:exp-regimes} details the evolution undergone by a hydrogen-oxygen detonation at different scenarios, with an initial pressure of 9 kPa and a temperature of 294 K. Specifically, Fig.\ \ref{reinitiation} shows the detonation transmission through 4 cylinders with a BR of 60\%. In this configuration, the detonation wave thickens during diffraction around the cylinders, then undergoes a Mach collision between the upper and lower branches of the leading shock at each obstacle's central axis. This collision results in the formation of a re-initiated detonation wave. Subsequently, the re-initiated detonation wave evolved back to its original structure before interacting with the obstacles. By adding one more cylinder to increase the BR to 75\%, the strength of the Mach shock formed during the collision between the upper and lower branches of the leading shock is insufficient to trigger re-initiation. Instead, a keystone-shaped hot spot emerges behind each Mach shock, and the flame gradually falls further behind the leading shock, as illustrated in Fig. \ref{hotspot}. By comparing Figures\ \ref{hotspot} and \ref{noinitiation}, one can see that the detonation transmission through either 5 or 10 cylinders with a BR of 75\% shows similar trends, indicting that once the detonation fails to initiate at the first Mach reflection, the fast flame cannot sustain.

\begin{comment}
\begin{figure}
	\begin{center}
		\subcaptionbox{$p_0$ = 10.3 kPa \label{noinitiation} }{\includegraphics[width=0.495\linewidth]{./figures/2H2-O2-1_5psi-Schlieren1C-dettripplepoint-5-8-12-18nb.jpg}}
		\subcaptionbox{$p_0$ = 12.1 kPa \label{hotspot} }{\includegraphics[width=0.495\linewidth]{./figures/2H2-O2-1_75psiT6-Schlieren1C-hotspot-6-9-12-15nb.jpg}}
		\subcaptionbox{$p_0$ = 12.4 kPa\label{reiniation} }{\includegraphics[width=0.495\linewidth]{./figures/2H2-O2-1_8psi-Schlieren1C-dettripplepoint-7-9-11-13-17nb.jpg}}
		\subcaptionbox{$p_0$ = 13.8 kPa\label{det} }{\includegraphics[width=0.495\linewidth]{./figures/2H2-O2-2_00psi-Schlieren1C-detsucess-5-7-9-11nb.jpg}}
	\end{center}
	%\captionsetup{labelformat=empty,labelsep=none}
	\caption{ Composite schlieren images illustrating (a) complete quenching, (b) hot-spot ignition behind the Mach shock (c) re-initiation from Mach reflection and (d) successfully transmission for stoichiometric hydrogen-oxygen detonation passing a cylinder with a diameter of 152.5 mm and 75\% blockage ratio at initial temperature 294 K.}
	\label{fig:exp-regimes}
\end{figure}
\end{comment}

\begin{figure}
	\begin{center}
		\includegraphics[width=0.54\linewidth]{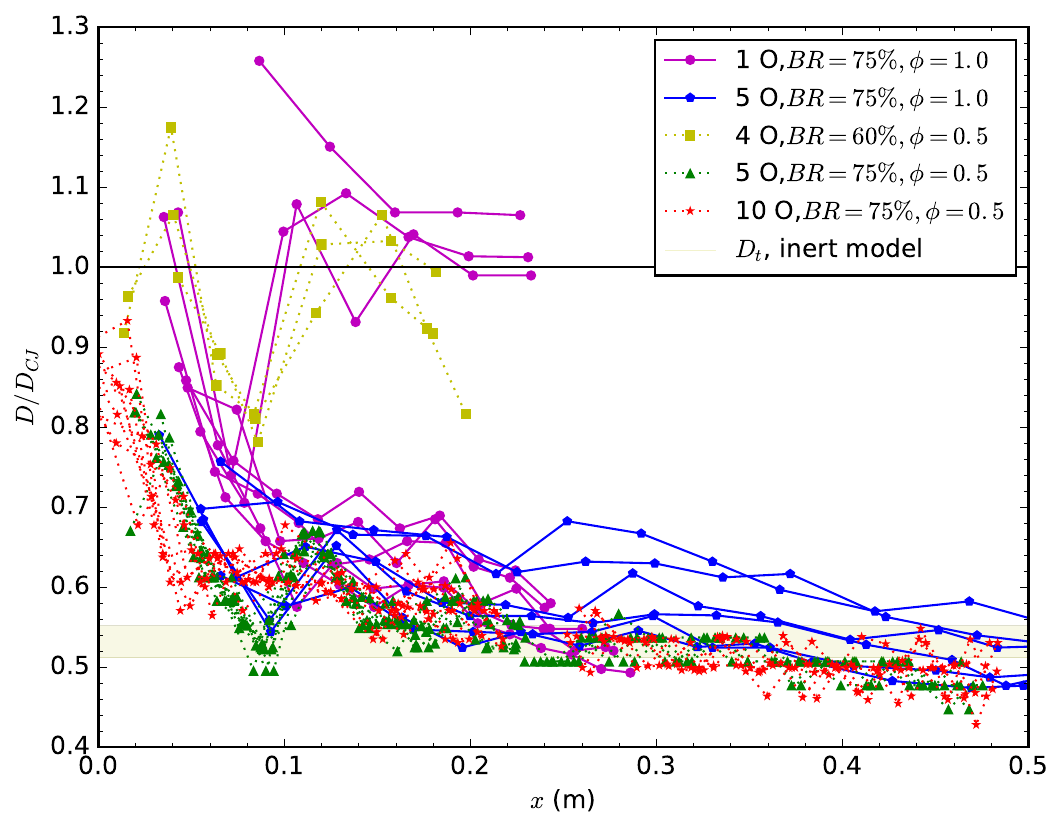}
	\end{center}
	%\captionsetup{labelformat=empty,labelsep=none}
	\caption{Shock speed profiles along the horizontal line at the center of the cylinders. The solid and dashed lines are for hydrogen-oxygen mixtures at $\phi=0.5$ in the current study and $\phi=1.0$ from Maley \cite{logan2015shock}, respectively. The $O$ shows the amount of obstacles, $x = 0$ represents the far right end of the obstacles. The measurement uncertainty is 149 m/s.}
	\label{fig:exp-comb}
\end{figure}

The velocities of the leading shock measured along the central axis of the obstacles at various initial pressures are presented in Fig.\ \ref{fig:exp-comb}. Additionally, the shock speeds for stoichiometric hydrogen-oxygen detonations transmission through obstacles from Maley \cite{logan2015shock} are included for comparison. As shown in the velocity plot, there are two distinct regimes in hydrogen detonation after interaction with cylinders: successful detonation transmission occurring immediately after the passage of the detonation wave through the obstacles, or complete quenching. For successful transmission, the leading shock speeds decrease to as low as 70\% CJ detonation speed ($D_{CJ}$) for $\phi=1.0$ and 80\% $D_{CJ}$ for $\phi=0.5$ mixture before being re-amplified by the Mach reflection to speeds greater than $D_{CJ}$. As shown in Fig.\ \ref{fig:ti-D}, these speeds correspond to the highest velocities of the cross-over regimes of the hydrogen mixtures. Although the Mach reflection at the centerline briefly amplifies the leading shock speed to at most 72\% $D_{CJ}$, the amplification was insufficient to re-initiate a detonation wave. Consequently, the shock speeds continues to decay after the Mach reflection.

%\section{Chemical kinetic ignition regimes}

\section{Detonation transmission regimes}
\label{subsec:3}

The experimental results from Maley \cite{logan2015shock} and the current study clearly indicate the presence of two distinct transmission outcomes for hydrogen-oxygen mixtures: successful transmission and complete quenching. Within a very limited range of pressures, detonation re-initiation and hot-spot ignition from Mach reflection are observed. In contrast to hydrocarbon mixtures, where failed detonation transmission typically leads to the presence of self-sustained fast flames, the behavior of hydrogen-oxygen mixtures differs. In this section, we will employ gas-dynamic and chemical kinetics analyses to explore and predict the different transmission regimes.

\subsection{Criterion for detonation transmission}

It was observed in Maley's experiments \cite{logan2015shock} that the critical pressure range for the successful transmission of a stoichiometric hydrogen-oxygen mixture through a single obstacle with a diameter of 152.5 mm and a blockage ratio of 0.75 lies between 13.8 kPa and 15.2 kPa. At these pressures, the equivalent inter-cylinder distance $b$ divided by a characteristic detonation cell size is $b/\lambda=4.5\pm3$, based on the available cell sizes in this mixture compiled by Shepherd \cite{kaneshige1997detonation}.  This limit corresponds to the commonly accepted limits for detonation diffraction from a channel to open space, which typically ranges between 3 and 7 \cite{radulescu2011mechanism, zangene2022critical}. 

Motivated by this observation, we addressed in more detail the diffraction process on cylinders with a high BR. The aim was to determine if both problems are indeed equivalent.  The question we asked is whether the curvature imposed by the diffraction process on the axis was the same for both geometries.  This falls within the curvature-induced failure phenomenology validated in our recent studies of detonation diffraction at abrupt area changes \cite{zangene2022critical, radulescu2021self}.  

Similar to Radulescu et al.\ \cite{radulescu2021self}, we set up to determine the curvature distribution on the leading shock induced by diffraction around a cylinder and compare it with that obtained at an abrupt area change. By adopting Whitham's geometric shock dynamics analysis \cite{whitham1974linear}, we can obtain the dynamics of a shock at any given time over a convex curved wall with the shape of a quarter circle. Assuming the gas in front of the shock is quiescent, the characteristic relations can be utilized to establish that both the shock deflection angle and the Mach number are constant along the straight lines intersecting the cylinder at different locations, defined by $\frac{dy}{dx} = \tan(\theta - \mu)$, where $\theta$ represents the shock deflection angle and $\tan\mu = \frac{Ac}{M}$, with $A$, $c$ and $M$ denoting the area of the ray tube, the speed of the nonlinear disturbance wave on the shock, and the shock Mach number, respectively. These lines propagate constant ray inclination angles and shock Mach numbers evaluated at the wall.  Our treatment of this problem provides the shock shape in closed form in the strong shock limit.  The coordinates of the shock are given in parametric form by
\begin{align}
\frac{X}{S_w t-r\sin(\frac{S_w t}{r M_0})} &= \exp(\frac{\theta}{\sqrt{n}}) (\cos\theta-\frac{\sin\theta}{\sqrt{n}}),\\
\frac{Y}{S_w t+\sqrt{n}(r-r\cos(\frac{S_w t}{r M_0}))} &= \exp(\frac{\theta}{\sqrt{n}}) (\sin\theta+\frac{\cos\theta}{\sqrt{n}}),
\end{align}
in the rectangular coordinate system originating from the edge of the circle following the shock. Here, $S_w$ is the normal speed of the shock with respect to the medium at rest, $r$ is the radius of the cylinder and $M_0$ is the initial shock Mach number. The shock deflection at the wall can be expressed as $\theta_w = \frac{S_w t}{r M_0}$. Additionally, $n$ is an arbitrary quantity defined as $n = -\frac{S_w^2 k}{\dot{S}w}$, where $k$ corresponds to the local curvature of the shock given by $k= \frac{X{_\theta} Y_{\theta \theta} - Y_{\theta} X_{\theta \theta}}{(X_{\theta}^2+Y_{\theta}^2)^\frac{3}{2}}$.

If the half-width of the channel is $b/2$, the time it takes for the initial transverse wave signal along the shock to reach $Y = r+b/2-rcos\theta_w$ can be calculated as $\frac{\sqrt{n}b}{2S_w}$, assuming $\theta_w$ is 0. At this specific time, the value of $k$ is determined using the same criteria as in Radulescu et al. \cite{radulescu2021self},
\begin{align}
k = \frac{\sqrt{n}}{n+1}\frac{2}{b}.
\end{align}
This shows that both the problems of detonation diffraction at an abrupt area change and over a circular wall provide the same curvature.  

Following Radulescu et al. \cite{radulescu2021self}, the critical channel height for successful detonation transmission can be determined by utilizing the critical curvature derived by He and Clavin \cite{he1994direct} for a quasi-steady square wave detonation, where the induction time is characterized by an exponential sensitivity to temperature. Since our problem of detonation diffraction passing through two cylinders spaced by $b$ was shown to be equivalent, the same model applies, yielding
\begin{equation}
\frac{b_*}{\Delta_i} = \frac{2\sqrt{n}}{n+1}\frac{8e}{(1-\gamma^{-2})} \left( \frac{Ea}{RT_N} \right),
\end{equation}
where $\Delta_i$ is the induction zone length of the CJ detonation, $e$ is the Euler's number, $\frac{Ea}{RT_n}$ is the activation energy non-dimensionalized by the temperature behind the shock of the CJ detonation.  The required parameters were determined through calculations using the constant volume homogeneous reactor approach provided by Cantera thermal chemical tools and the Li mechanism \cite{li2004updated}. The exponent $n$ in the equation relies on the chosen model for shock evolution. Therefore, the range of values for $n$ obtained from the widely employed models, such as Whitham \cite{whitham1974linear}, Wescott et al. \cite{wescott2004self}, and Radulescu et al. \cite{radulescu2021self}, was utilized to calculate $\frac{b_*}{\Delta_i}$, as presented in Table \ref{table_model_diff}. The model predicts the critical value within less than a factor of 2, indicating reasonably good agreement. Note that our treatment neglects the change in detonation overdrive at the throat of the cylinders and likewise neglects the Mach reflection of the detonation on the converging section, which may enhance the transmission process. This might explain the over prediction of the model. A comprehensive investigation of these effects is beyond the scope of the present study and warrants further exploration in future research.

%but this overdrive is expected to be eroded by expansion waves originating from the back  as the detonation is exposed to simultaneous lateral diffraction.  Accounting for detonation overdrive was found to be important by Desbordes et al.\ \cite{desbordes1988transmission} in correlating diffraction limits.  Accounting for this effect will lower the model prediction in the direction of the experiment.  

It is also worth mentioning that the experiments demonstrate that the critical pressure for detonation re-initiation from a Mach reflection is close to that of the direct successful transmission for hydrogen detonations passing through small openings. This scenario specifically refers to obstacles with a large BR and the obstacle radius is much larger than the hole size, effectively recovering the detonation diffraction limit. We can use the current model to predict detonation transmission in such cases. However, for the cases with low BR where the obstacle radius is close to the hole size, it appears to remain a transmission problem. For example, as shown in Fig.\ \ref{fig:exp-comb}, the detonation re-initiation occurs at $\frac{b}{\Delta_i} = 26$ with $b$ being 16.2 mm, which is much smaller than the critical value for successful transmission. Notably, the cell size of the detonation wave under the condition in Fig.\ \ref{fig:exp-comb} is $15 \sim 20$ mm. This seems to align well with previous studies for the detonation limit problem, where $b_*$ is empirically observed to be approximately 1 cell size \cite{radulescu2011mechanism,ciccarelli2008flame}.

\begin{table}
	\begin{center}
		\caption{Summary of the model prediction for hydrogen-oxygen detonation transmission at T=294 K and $p_0$ = 13.8 - 15.2 kPa. }
		\def~{\hphantom{0}}
		\begin{tabular}{cc}
			\hline
			%\multicolumn{3}{|c|}{OneTwoThree}
			$b_*/\Delta_i$ (model)  & $b_*/\Delta_i$ (exp)\\
			\hline
			190-217 & 115-128\\
			\hline
		\end{tabular}
		\label{table_model_diff}
	\end{center}
\end{table}

%Radulescu and Maxwell\cite{radulescu2011mechanism} have found that the critical conditions for detonation re-initiation correspond approximately to a pore size equal to approximately 30 – 60 detonation induction lengths for acetylene-oxygen mixtures, which were also known to have the cross-over regimes. In the current, the $\frac{W_*}{\Delta_i}$ for detonation re-initiation was found to be 104 detonation induction lengths, which is close to the critical successful transmission. 

\subsection{Complete quenching}

\begin{figure}
	\begin{center}
		\includegraphics[width=0.57\linewidth]{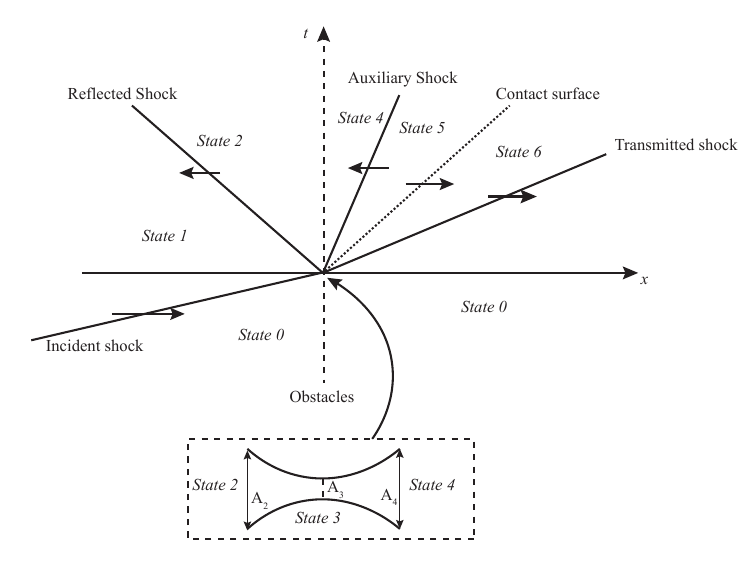}
	\end{center}
	%\captionsetup{labelformat=empty,labelsep=none}
	\caption{Space-time diagram illustrating the self-similar gasdynamic model for the interaction of a detonation with a row of cylinders.}
	\label{fig:InertModel}
\end{figure}

To analyze completely quenched detonation waves, we adopted the inert quasi-one-dimensional model proposed in \cite{radulescu2015chapman, wang2019models} to reconstruct the flow field and evaluate the shock speed. As depicted in Fig.\ \ref{fig:InertModel}, the passage of the detonation wave through the obstacles is presumed to generate a transmitted shock and a reflected shock. The detonation wave is represented by an inert shock characterized by a Mach number of $M_{CJ}$. The over-expanded sonic jet flow that emerges through the pores, exhibiting a series of shock diamonds in multidimensional scenarios, is modeled by an auxiliary shock. A contact surface is utilized to represent the jet head, which separates the gases initially positioned on the left side of the obstacles from those on the right. The flow from state 2 to state 4 is considered as isentropic converging-diverging nozzle flow, where the area ratio $\frac{A_3}{A_2} =\frac{A_3}{A_4} $ is equal to 1 minus the BR. Additionally, the flow at state 3 is assumed to be choked. The shock transitions comply with the conventional shock jump relations. A two-gamma approximation is employed, with the unburned isentropic exponent assigned to state 0 and state 6, while the CJ detonation isentropic exponent is assigned to states 2 to 5.

%The model predicts a transmitted shock speed of approximately $51-56\%$ of $D_{CJ}$ for completely quenched hydrogen-oxygen detonations at $\phi=1.0$ and $\phi=0.5$ interacting with cylinders with a blockage ratio of 0.75. As shown in Fig.\ \ref{fig:exp-comb}, although the Mach reflection between the upper and lower branch of the leading shock passing the cylinder briefly boosted the shock speed to at most 72\% $D_{CJ}$, the leading shock speed eventually decayed to approximately 50\% $D_{CJ}$. Therefore, the model appears to agree well with the experimental results. 

The transmitted shock speeds of completely quenched hydrogen-oxygen detonations at $\phi=1.0$ and $\phi=0.5$, interacting with cylinders with a blockage ratio of 0.75, were predicted and compared with experimental data. As shown in Fig.\ \ref{fig:exp-comb}, the model's predicted transmitted shock speeds exhibit good agreement with the experimental results once the shock has traveled approximately 0.1 m from the end of the obstacles, at which the Mach shock has formed. For a completely quenched hydrogen-oxygen detonation wave, the transmitted shock speed is approximately $50 \sim 60\%$ of $D_{CJ}$. Although the first Mach reflection briefly boosted the shock speed to at most 72\% $D_{CJ}$, it is proved insufficient to re-initiate a detonation wave. Consequently, the leading shock speed eventually decayed to the predicted value.

\begin{figure}
	\begin{center}
		\includegraphics[width=0.53\linewidth]{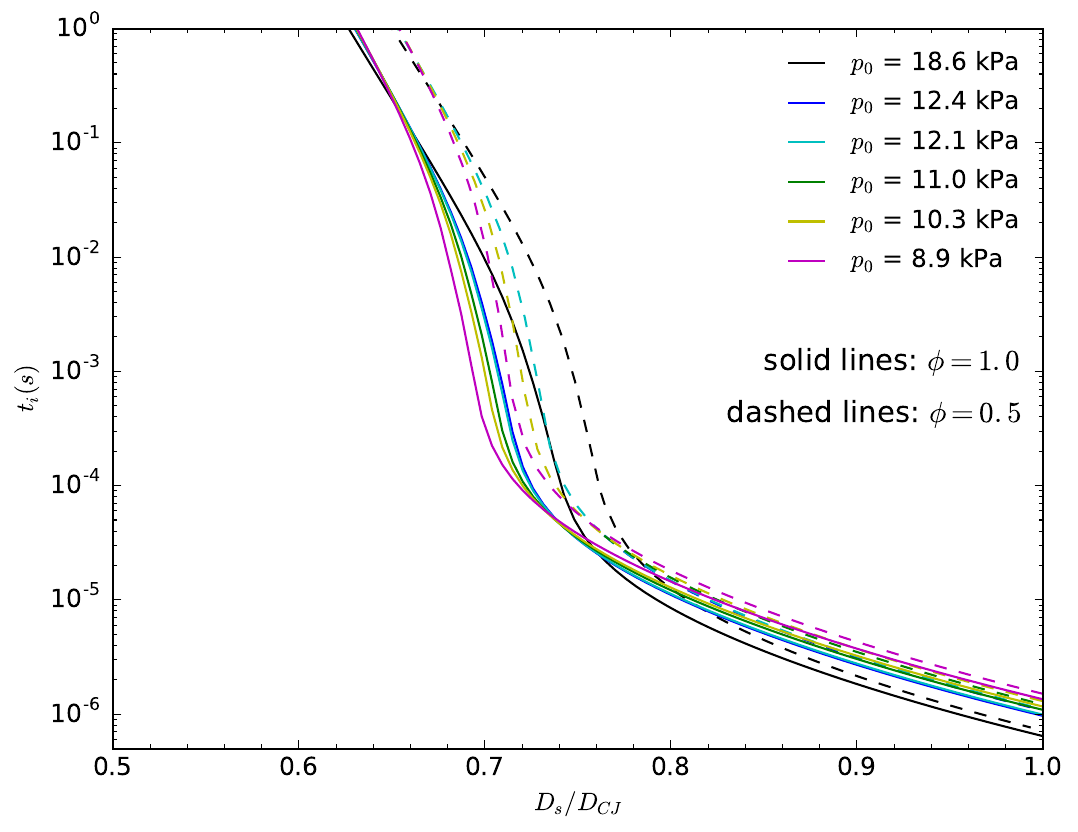}
	\end{center}
	%\captionsetup{labelformat=empty,labelsep=none}
	\caption{The ignition delay time vs. shock speed at various initial pressures and equivalence ratios illustrating the chain branching nature of hydrogen.}
	\label{fig:ti-D}
\end{figure}

The ignition mechanism was then investigated by calculating the ignition delay time for the gas in the post-Mach shock regime. This was achieved using constant volume homogeneous reactor calculations implemented with Cantera thermal chemical tools and the Li mechanism \cite{li2004updated}. In Figures \ref{fig:exp-comb} and \ref{fig:ti-D}, it can be observed that immediately after traversing the cylinders, the decoupled shocks rapidly enter the cross-over regime characterized by shock speeds ranging from 0.65 to 0.8 $D_{CJ}$, a behavior specific to hydrogen mixtures. Within this regime, the ignition delay time spans from $10^{-5}$ to $10^{-1}$ s, and even a slight variation in shock speed can significantly impact the ignition process. This probably explains the very narrow regimes of detonation re-initiation and hot-spot ignition. As the Mach reflection lacks sufficient strength to produce an ignition delay time on the order of the obstacle diameter divided by the detonation velocity (approximately $5 \times 10^{-5}$ s), which is several orders of magnitude smaller, the limited reactivity of the auto-ignited hot spots result in quenching. This suggests that attenuating hydrogen mixtures may be comparatively easier than attenuating other hydrocarbon mixtures due to their chain-branching characteristics. In the case of a completely quenched detonation, the Mach shock speed is at most 72\% $D_{CJ}$. Auto-ignition is not feasible in this regime. Given the satisfactory agreement between the predictions and experiments, the inert-quasi-one-dimensional model and chemical kinetic analysis can be utilized to determine the critical blockage ratio at a desired initial pressure.

\subsection{Significance of hydrogen to other hydrocarbon mixtures in detonation re-initiation}

The experimental results clearly demonstrate that when the oxy-hydrogen detonation fails, the state behind the leading shock falls into or below the cross-over regime. Maley suggested \cite{logan2015shock} that the hydrodynamic jet is responsible for the detonation re-initiation, and this is influenced by the isentropic exponent ($\gamma$). For the hydrogen mixtures, the large $\gamma$ prevents the formation of strong jets behind periodically formed Mach shocks, providing local enhancement of the reactivity through turbulent mixing. Thus, there is no mechanism for the failed detonation to re-amplify. However, as pointed out by previous studies \cite{logan2015shock,saif2017chapman} with the same BR, quenched hydrocarbon-oxygen detonations, including stoichiometric mixture of CH$_4$, C$_2$H$_4$, C$_2$H$_6$ and C$_3$H$_8$, can re-amplify after traversing a substantial distance past obstacles, with the turbulent reaction zone structure closely coupled with the leading shock. Only the detonation waves of the 2C$_2$H$_2$/5O$_2$/21Ar shows a similar trend as the oxy-hydrogen detonation. This has been demonstrated to correlate well with the mixture sensitivity parameter $\chi$. The high value of $\chi$ appears to lead to the re-initiation in hydrocarbons.

\begin{figure}
	\begin{center}
		\subcaptionbox{\label{ti-comp} }{\includegraphics[width=0.325\linewidth]{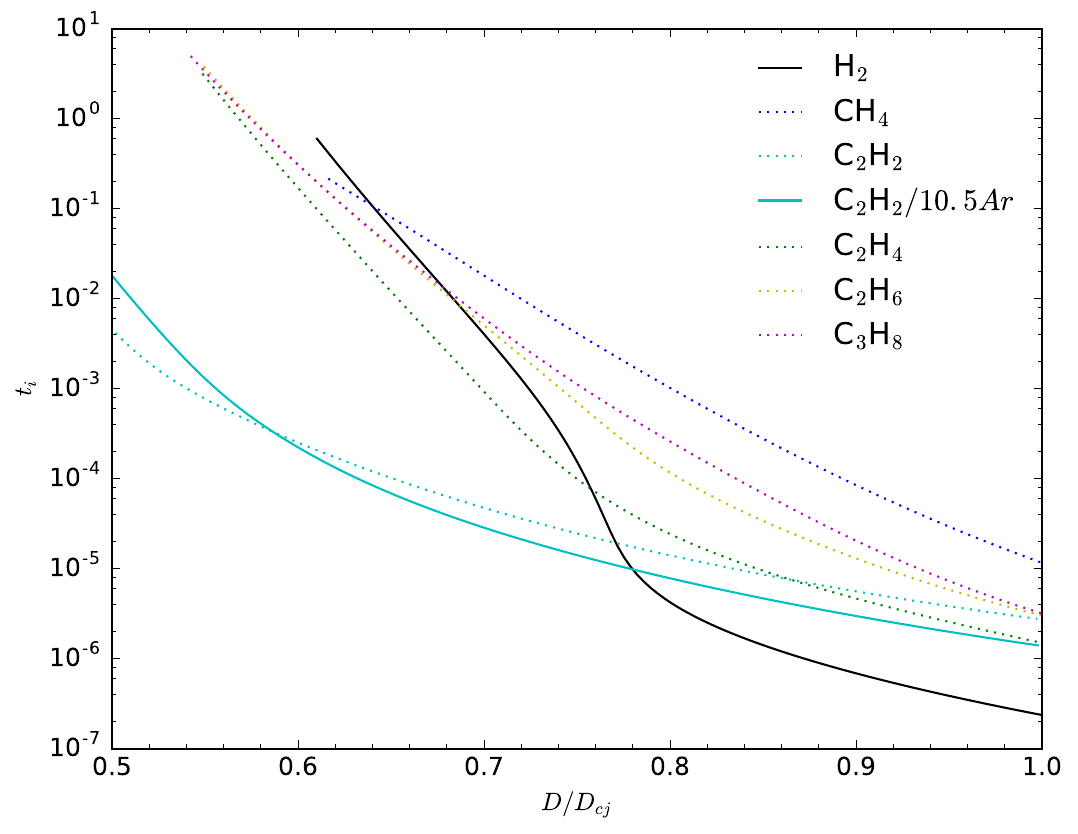}}
		\subcaptionbox{\label{chi-comp} }{\includegraphics[width=0.325\linewidth]{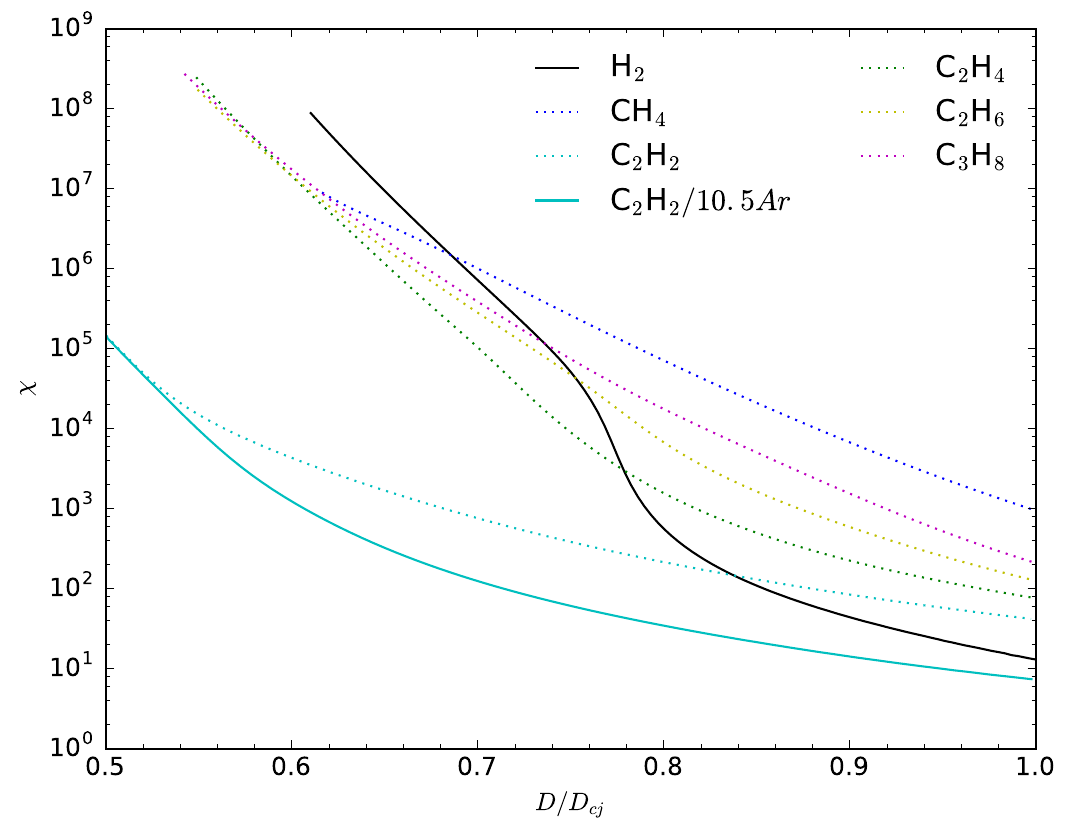}}
		\subcaptionbox{ \label{gamma-comp} }{\includegraphics[width=0.325\linewidth]{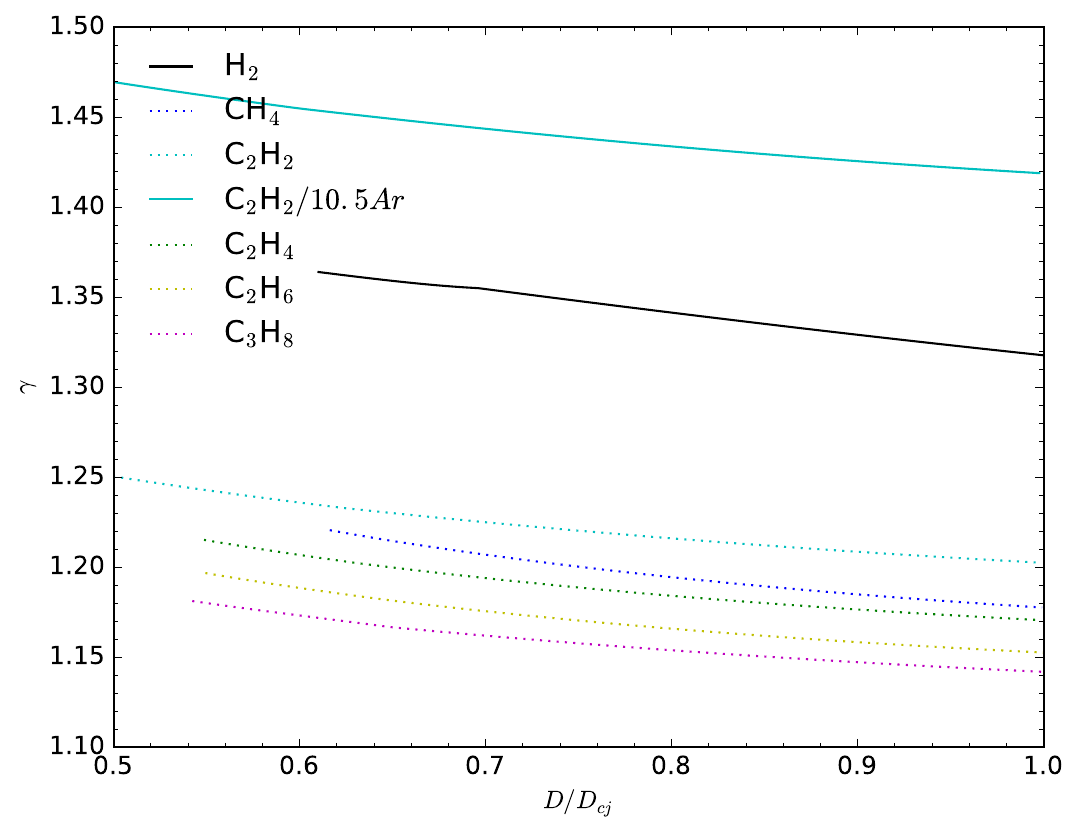}}
		%\subcaptionbox{$p_0$ = 13.8 kPa\label{det} }{\includegraphics[width=0.495\linewidth]{./figures/2H2-O2-2_00psi-Schlieren1C-detsucess-5-7-9-11nb.jpg}}
	\end{center}
	%\captionsetup{labelformat=empty,labelsep=none}
	\caption{ Comparison of (a) ignition delay time $t_i$, (b) mixture sensitivity parameter $\chi$ and (c) isentropic exponent $\gamma$ as a function of leading shock speed for different hydrocarbon and hydrogen mixtures at initial pressures corresponding to the critical re-initiation pressures.}
	\label{fig:ti-gamma-comp}
\end{figure}

To further explore the role of hydrodynamic instability and chemical kinetics, we conducted experiments in 2C$_2$H$_2$/5O$_2$ and observed the detonation re-initiation after a long distance past the obstacles. The fast flame was found to be able re-amplify to a detonation with a leading shock speed of 55\% $D_{CJ}$. We then calculated the ignition delay time $t_i$, mixture sensitivity parameter $\chi$, and isentropic exponent $\gamma$ for the aforementioned hydrogen and hydrocarbon mixtures at initial pressures corresponding to the critical re-initiation pressures using the Blanquart et al. mechanism (V2.3) \cite{Blanquart2018}. The available data from previous studies and the current experimental results show that the critical transmitted shock speed is in the range of $50 \sim 75\%$ of $D_{CJ}$ for all the mixtures considered, specifically 60\% $D_{CJ}$ for the 2C$_2$H$_2$/5O$_2$/21Ar mixture. Figure\ \ref{fig:ti-gamma-comp} compares the calculated $t_i$, $\chi$ and $\gamma$ at such leading shock speeds. It can be observed that the re-initiation of the 2C$_2$H$_2$/5O$_2$ detonations is a result of auto-ignition because of the small $t_i$, whereas the 2C$_2$H$_2$/5O$_2$/21Ar fast flame falls into the chain-branching cross-over regime, similar to the hydrogen mixture, and fails to re-amplify. In other hydrocarbon mixtures, both the $t_i$ and $\chi$ are high, similar to that of the hydrogen mixtures. This means that one cannot solely attribute the re-initiation in these mixtures to auto-ignition and purely chemical kinetics. Instead, as shown in Fig.\ \ref{gamma-comp}, the $\gamma$ effect appears to have a better correlation for the re-initiation. As is suggested by \cite{mach2011mach, lau2021viscous}, when $\gamma$ exceeds 1.32, the hydrodynamic jet is not strong enough to form the shock bifurcation on the front of the Mach shock. This $\gamma$ criteria seems to be compatible with all mixtures, suggesting that the hydrodynamic instability is important. 

\section{Conclusions}
\label{subsec:4}
In summary, this study revisited the issue of detonation transmission through a bank of cylinders in hydrogen mixtures and identified two distinct regimes: successful transmission and complete quenching. In a narrow range of pressures, re-initiation from shock reflection and hot spot ignition behind the first Mach shock can occur in failed detonations. The critical channel height for successful detonation transmission was found to be approximately 2 to 7 times the cell size, which aligns with the widely accepted limits for diffraction in critical slot configurations. Whitham's geometrical shock dynamics analysis, following the procedure proposed by Radulescu et al. \cite{radulescu2021self} for detonation diffraction around sharp corners, was utilized to evaluate the critical detonation transmission criteria. The model demonstrated good agreement with experimental results.

When the detonation does not survive the diffraction process, the lead shock is found to decay to approximately 50\% to 60\% of $D_{CJ}$. Although the Mach reflection can boost the leading shock speed to as high as 72\%  $D_{CJ}$, the temperature still falls into the chain-branching cross-over ignition limit for hydrogen ignition, effectively quenching the ignition process.  The cross-over occurs for shocks propagating between approximately 65\% to 80\% of $D_{CJ}$. In this regime, the very long $t_i$ suppresses auto-ignition, and the high $\gamma$ prevents further convective mixing for re-initiation. We propose and validate a simple gasdynamic model for evaluating the strength of the transmitted shock in terms of blockage ratio. Together with the Mach reflection and chain-branching auto-ignition limit, this provides simple guidelines for arresting detonations in hydrogen mixtures. Further experiments should be devoted towards studying these conditions.   

\section{Acknowledgments}
\label{sec:5}
The authors thank Zhe (Rita) Liang from Canadian Nuclear Laboratories for the sponsorship and useful discussions.

\bibliographystyle{ieeetran}
%\bibliography{references.bib}
\bibliography{references} 

% Generated by IEEEtran.bst, version: 1.14 (2015/08/26)
\begin{thebibliography}{10}
\providecommand{\url}[1]{#1}
\csname url@samestyle\endcsname
\providecommand{\newblock}{\relax}
\providecommand{\bibinfo}[2]{#2}
\providecommand{\BIBentrySTDinterwordspacing}{\spaceskip=0pt\relax}
\providecommand{\BIBentryALTinterwordstretchfactor}{4}
\providecommand{\BIBentryALTinterwordspacing}{\spaceskip=\fontdimen2\font plus
\BIBentryALTinterwordstretchfactor\fontdimen3\font minus
  \fontdimen4\font\relax}
\providecommand{\BIBforeignlanguage}[2]{{%
\expandafter\ifx\csname l@#1\endcsname\relax
\typeout{** WARNING: IEEEtran.bst: No hyphenation pattern has been}%
\typeout{** loaded for the language `#1'. Using the pattern for}%
\typeout{** the default language instead.}%
\else
\language=\csname l@#1\endcsname
\fi
#2}}
\providecommand{\BIBdecl}{\relax}
\BIBdecl

\bibitem{radulescu2011mechanism}
M.~I. Radulescu and B.~M. Maxwell, ``The mechanism of detonation attenuation by
  a porous medium and its subsequent re-initiation,'' \emph{Journal of Fluid
  Mechanics}, vol. 667, pp. 96--134, 2011.

\bibitem{logan2015shock}
L.~Maley, ``On shock reflections in fast flames,'' Master's thesis,
  Universit{\'e} d'Ottawa/University of Ottawa, 2015.

\bibitem{saif2017chapman}
M.~Saif, W.~Wang, A.~Pekalski, M.~Levin, and M.~I. Radulescu,
  ``Chapman--{J}ouguet deflagrations and their transition to detonation,''
  \emph{Proceedings of the Combustion Institute}, vol.~36, no.~2, pp.
  2771--2779, 2017.

\bibitem{yang2022experimental}
T.~Yang, Q.~He, J.~Ning, and J.~Li, ``Experimental and numerical studies on
  detonation failure and re-initiation behind a half-cylinder,''
  \emph{International Journal of Hydrogen Energy}, vol.~47, no.~25, pp.
  12\,711--12\,725, 2022.

\bibitem{kaneshige1997detonation}
M.~Kaneshige and J.~E. Shepherd, ``Detonation database,'' 1997.

\bibitem{zangene2022critical}
F.~Zangene, Q.~Xiao, and M.~Radulescu, ``Critical diffraction of irregular
  structure detonations and their predictability from experimentally obtained
  {D}- $\kappa$ data,'' \emph{Proceedings of the Combustion Institute}, 2022.

\bibitem{radulescu2021self}
M.~I. Radulescu, R.~M{\'e}vel, Q.~Xiao, and S.~Gallier, ``On the
  self-similarity of diffracting gaseous detonations and the critical channel
  width problem,'' \emph{Physics of Fluids}, vol.~33, no.~6, p. 066106, 2021.

\bibitem{whitham1974linear}
G.~Whitham, \emph{Linear and nonlinear waves}.\hskip 1em plus 0.5em minus
  0.4em\relax New York, Wiley-Interscience, 1974.

\bibitem{he1994direct}
L.~He and P.~Clavin, ``On the direct initiation of gaseous detonations by an
  energy source,'' \emph{Journal of Fluid Mechanics}, vol. 277, pp. 227--248,
  1994.

\bibitem{li2004updated}
J.~Li, Z.~Zhao, A.~Kazakov, and F.~L. Dryer, ``An updated comprehensive kinetic
  model of hydrogen combustion,'' \emph{International Journal of Chemical
  Kinetics}, vol.~36, no.~10, pp. 566--575, 2004.

\bibitem{wescott2004self}
B.~Wescott, D.~S. Stewart, and J.~Bdzil, ``On self-similarity of detonation
  diffraction,'' \emph{Physics of Fluids}, vol.~16, no.~2, pp. 373--384, 2004.

\bibitem{ciccarelli2008flame}
G.~Ciccarelli and S.~Dorofeev, ``Flame acceleration and transition to
  detonation in ducts,'' \emph{Progress in energy and combustion science},
  vol.~34, no.~4, pp. 499--550, 2008.

\bibitem{radulescu2015chapman}
M.~I. Radulescu, W.~Wang, M.~Saif, M.~Levin, and A.~Pekalski, ``On {C}hapman
  {J}ouguet deflagrations,'' in \emph{In Proceedings of 25th International
  Colloquium on the Dynamics of Explosions and Reactive Systems}, 2015.

\bibitem{wang2019models}
W.~Wang, ``Models of {CJ} deflagrations and their transition to detonations
  from the interaction of a detonation wave with a perforated plate,'' Master's
  thesis, Universit{\'e} d'Ottawa/University of Ottawa, 2019.

\bibitem{Blanquart2018}
G.~Blanquart, \emph{CaltechMech v2.3}.\hskip 1em plus 0.5em minus 0.4em\relax
  Available at http://theforce.caltech.edu/resources/, 2015.

\bibitem{mach2011mach}
P.~Mach and M.~Radulescu, ``Mach reflection bifurcations as a mechanism of cell
  multiplication in gaseous detonations,'' \emph{Proceedings of the Combustion
  Institute}, vol.~33, no.~2, pp. 2279--2285, 2011.

\bibitem{lau2021viscous}
S.-M. Lau-Chapdelaine, Q.~Xiao, and M.~Radulescu, ``Viscous jetting and mach
  stem bifurcation in shock reflections: experiments and simulations,''
  \emph{Journal of Fluid Mechanics}, vol. 908, p. A18, 2021.

\end{thebibliography}


\begin{thebibliography}{99}
	
	\bibitem[Radulescu et al. (2011)]
	{Radulescu2011}
	Radulescu, Matei I., and Maxwell, B.: The mechanism of detonation attenuation by a porous medium and its subsequent re-initiation. Journal of Fluid Mechanics 667 (2011): 96-134.
	
	\bibitem[Maley (2015)] 
	{Maley2015}
	Maley, L.: On Shock Reflections in Fast Flames. Diss. Université d'Ottawa/University of Ottawa, 2015.
	
	\bibitem[Yang et al. (2022)]
	{Yang2022}
	Yang, T., He, Q., Ning, J., \& Li, J.: Experimental and numerical studies on detonation failure and re-initiation behind a half-cylinder. International Journal of Hydrogen Energy, 47(25), 12711-12725 (2022).
	
	\bibitem[Li et al. (2022)]
	{Li2004}
	Li, J., Zhao, Z., Kazakov, A., \& Dryer, F. L.: An updated comprehensive kinetic model of hydrogen combustion. International journal of chemical kinetics, 36(10), 566-575(2004).
	
	\bibitem[YangH et al. (2022)]
	{YangH2022}
	Yang, H. X., Rakotoarison W., Sow A., Liang Z., Radulescu M. I.: The pressure dynamics from head-on reflections of detonations and high-speed deflagrations in lean hydrogen mixtures, {Proceedings of Combustion Institute-Canadian Section}, Ottawa, Canada, 15th -- 19th May, 2022.
	
	\bibitem[Oppenheim et al. (1968)]
	{Oppenheim1968}
	Oppenheim, A. K., Smolen, J. J., \& Zajac, L. J. (1968). Vector polar method for the analysis of wave intersections. Combustion and Flame, 12(1), 63-76.
	
\end{thebibliography}
\begin{comment}

\end{comment}

\end{document}